\documentclass[lettersize,journal]{IEEEtran}
\usepackage{amsmath,amsfonts}
\usepackage{algorithmic}
\usepackage{algorithm}
\usepackage{array}
\usepackage[caption=false,font=normalsize,labelfont=sf,textfont=sf]{subfig}
\usepackage{textcomp}
\usepackage{stfloats}
\usepackage{url}
\usepackage{verbatim}
\usepackage{graphicx}
\usepackage{cite}
\begin{document}

\title{SecureBERT and L{\huge LAMA} 2 Empowered Control\\ Area Network Intrusion Detection and Classification}

 


\author{
 Xuemei Li,~\IEEEmembership{Student Member,~IEEE},
Huirong Fu,~\IEEEmembership{Member,~IEEE}
}





\maketitle

\begin{abstract}
Numerous studies have proved their effective strength in detecting Control Area Network (CAN) attacks. In the realm of understanding the human semantic space, transformer- based models have demonstrated remarkable effectiveness. Leveraging pre-trained transformers has become a common strategy in various language-related tasks, enabling these models to grasp human semantics more comprehensively. To delve into the adaptability evaluation on pre-trained models for CAN intrusion detection, we have developed two distinct models: CAN-SecureBERT and CAN-L{\scriptsize LAMA}2. Notably, our CAN-L{\scriptsize LAMA}2  model surpasses the state-of-the-art models by achieving an exceptional performance 0.999993 in terms of balanced accuracy, precision detection rate, F1 score, and a remarkably low false alarm rate of 3.10e-6. Impressively, the false alarm rate is 52 times smaller than that of the leading model, MTH-IDS (Multitiered Hybrid Intrusion Detection System). Our study underscores the promise of employing a Large Language Model as the foundational model, while incorporating adapters for other cybersecurity-related tasks and maintaining the model’s inherent language-related capabilities.
\end{abstract}


\section{Introduction}
\IEEEPARstart{V}{ehicle} CAN constitutes a standardized communication protocol extensively employed within the automotive industry. It serves as a primary communication interface connecting the vehicle gateway, Electronic Control Units (ECUs), and various other control components integral to vehicular operations. For an Intrusion Detection System (IDS), monitoring the CAN messages between ECUs to enable a comprehensive understanding of the communication processes and the identification of anomalous messages are essential. The United Nations Regulation WP.29 R155 [1] has articulated a new mandate, rendering it obligatory for all new vehicles manufactured within the European Union to adhere to this regulation, effective from July 2024. This regulation enforces all vehicles to possess the capability to detect and respond to potential cybersecurity attacks.  Moreover, there is a requisite for the systematic collection of log data to facilitate the identification of cyber attacks and to support forensic investigations. Therefore, an adaptable and multifaceted solution capable of meeting both demands becomes imperative.

Based on technology implementation, four primary methodologies for CAN intrusion detection have been identified,  visualized in Figure 1 [2]. These methodologies are proficient in detecting CAN intrusion attacks but do not possess the inherent capability to collect and analyze log data. The first is fingerprint-based, which involves the detection of anomalies through clock-based or voltage measurements. The second, performed at the message level, is parameter monitoring-based and comprises techniques such as frequency-based, whitelist-based, or remote frame-based analysis. The third operates at the data-flow level and is grounded in information theory, employing entropy analysis or hamming distance measurements. The fourth takes advantage of Machine Learning (ML) techniques at a functional level. ML methodologies have undergone extensive evaluation in the context of CAN intrusion detection, with a focus on Recurrent Neural Networks (RNNs), Deep Neural Networks (DNNs), and Artificial Neural Networks (ANNs). Additionally, ML models like Decision Tree (DT)-based or Hidden Markov Model (HMM)-based models have found application in this domain.  

\begin{figure*}[!t]
\centering
\includegraphics[scale=0.56]{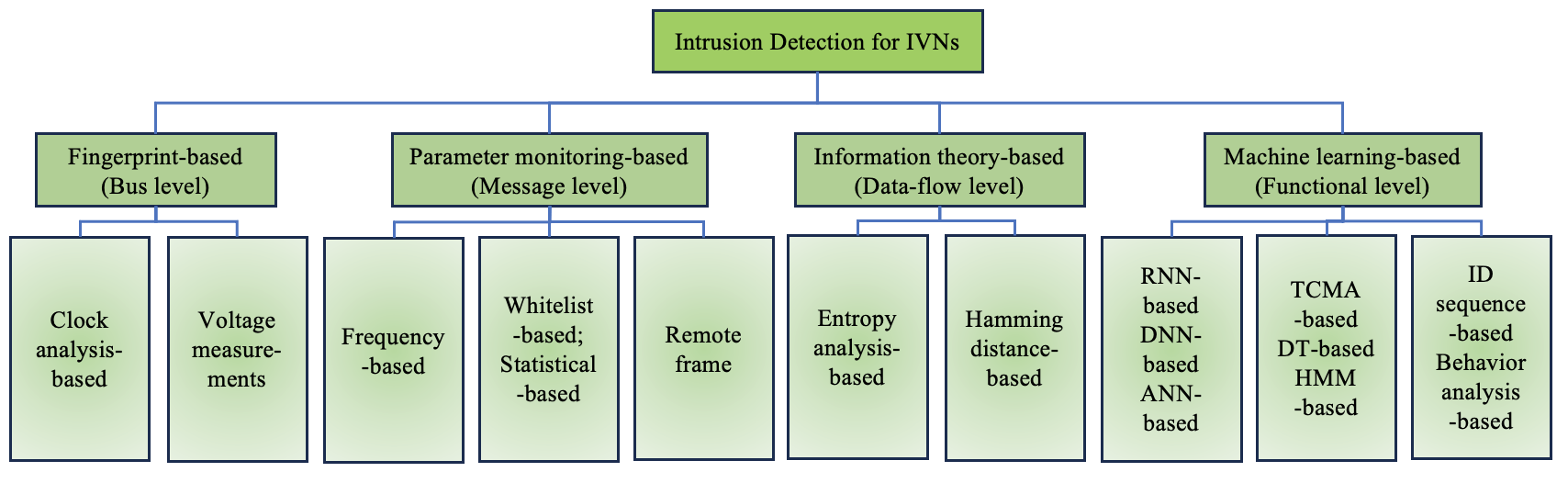}
\caption{In-Vehicle Networks (IVNs) IDS Taxonomy Based on Technology Implementation [2]}
\label{Fig1}
\end{figure*}

The primary limitation of existing implementations in CAN intrusion detection is their dependence on supplementary access, such as physical layer interactions, preprocessing of CAN messages, extraction of corresponding features, and creation of intricate rules and work flows within software, which leads to numerous additional requirements for effective CAN intrusion detection, while limiting the capacity to identify unknown and novel attacks. To address these constraints, Cho and Shin et al. [3] developed Viden, a novel approach that entails fingerprinting ECUs based on their corresponding voltage measurements. Viden acquires precise voltage measurements directly from the message transmitters and subsequently processes this data to construct accurate voltage profiles and fingerprints. The method is primarily applicable to physical layer attacks. Furthermore, frequency-based techniques adopt a different perspective by focusing on periodic traffic, rooted in the observation that frequencies tend to increase when malevolent adversaries engage in spoofing or Denial of Service (DoS) attacks [4].  On the other side,  CAN IDS using the traditional ML models such as DTs,  Random Forests (RFs), and Neural Networks, require preprocess CAN message and perform feature engineering to train the model in order to perform classification.

Transformer models have garnered recognition for the remarkable proficiency in grasping human semantics and handling various Natural Language Processing (NLP) tasks. They possess the capacity to accept textual input and generate output directly. Their remarkable capabilities have been well-demonstrated in the realms of computer vision and NLP tasks.  The first introduce of the transformer model can be attributed to Ashish et al. [5]. Its architecture comprises an encoder and decoder structure. This design replaced the traditional recurrent-based models' recurrent layer with a multi-head self-attention layer, leading to substantial improvements in performance, particularly in translation tasks.

There are a couple of motivations for our study to integrate transformer models. First, transformer models can be pre-trained on extensive unlabeled datasets. This pre-training can significantly enhance model performance during transfer learning [6]. This technique is commonly referred to as domain adaptation, which falls under transductive transfer learning. It is specialized to address challenges arising from a substantial volume of unlabeled source data for training and limited unlabeled target data [7].  This strategy has demonstrated its efficacy by achieving state-of-the-art results in various common NLP benchmarks, as evidenced in literature  [8]. The empirical robustness of this strategy extends across diverse domains, including computer vision, audio, and text processing tasks.

This paper endeavors to adapt pre-trained transformer-based models for the purpose of detecting CAN attacks. Leveraging the inherent capabilities of transformers has the potential to markedly enhance efficiency and streamline the architecture of an IDS for vehicular CAN attacks. Transformers offer a range of advantages, the first being their ability to learn intrinsic relationships through position encoding and a multi-head self-attention mechanism. The second advantage is that transformers can learn meaningful representations from CAN message logs without the need for extensive data preprocessing or feature extraction. 

In this study,  \textbf{ Our Primary Contributions} are summarized as follows: 
\begin{enumerate}
\item We propose CAN-C-BERT, CAN-SecureBERT and CAN-LLAMA2 models to detect and classify various CAN attacks by adapting pre-trained transformer-based models, namely BERT [9], SecureBERT [10],  and L{\scriptsize LAMA} 2 [11].
\item We design and demonstrate detailed architectures of the proposed three models and discuss how to train them to detect and classify CAN attacks. 
\item We develop and construct the proposed CAN-C-BERT, CAN-SecureBERT and CAN-LLAMA2 models by integrating pre-trained models with classification heads and training the proposed models with pre-balanced CAN dataset. 
\item We conduct empirical study of the proposed models and compare them with the state-of-the-art model MTH-IDS [12]. The experimental study shows that our proposed CAN-L{\scriptsize LAMA}2 model demonstrates the highest performance and CAN-SecureBERT secures the second-highest position.
\end{enumerate}

In particular, our work has unique innovations which yield significant results, including: 

\begin{enumerate}
\item{\textbf{Utilization of CAN Message Logs}}: Our proposed CAN-C-BERT, CAN-SecureBERT and CAN-L{\scriptsize LAMA}2 models possess the unique ability to directly employ CAN message logs for intrusion detection and attack classification. This approach eliminates the need for traditional data preprocessing, as the models can directly analyze the raw data. 
\item {\textbf{Superior Performance with Limited Data and Enhanced Generalization with Larger Data}}:  Our proposed CAN-C-BERT, CAN-SecureBERT and CAN-L{\scriptsize LAMA}2 models have achieved superior performance while training only 5\% of the data, while other state-of-art models ([13]-[21]) require a larger dataset to achieve a performance level similar to ours. Furthermore, our models trained with more extensive datasets exhibit improved generalization and outperform their counterparts trained with smaller dataset. 
\item{\textbf{Domain Knowledge}}: The incorporation of cybersecurity domain knowledge in CAN-SecureBERT does not directly contribute to the detection and classification of CAN attacks when compared with other pre-trained models. 
\item {\textbf{LoRA Utilization in CAN-L{\scriptsize LAMA}2}}: Leveraging the Low-Rank Adaption (LoRA) [22] technique for training CAN-L{\scriptsize LAMA}2 resulted in a modification of only 0.57\% of the model’s parameters. This demonstrates that the majority of the original L{\scriptsize LAMA} 2 model parameters remain unchanged. Consequently, the CAN-L{\scriptsize LAMA}2 model retains its versatility and can be employed for various language-related tasks. For instance, the Vehicle Security Operations Center (VSOC) team can capitalize on this model and adapt it for diverse downstream tasks by fine-tuning the pre-trained model and integrating adapter heads. 
\end{enumerate}

The remainder of this paper is organized as follows: Section II presents an extensive literature review. Section III conducts a comparative analysis of the model architectures, highlighting the distinctions between CAN-C-BERT, CAN SecureBERT, and CAN-L{\scriptsize LAMA}2. Section IV outlines the key techniques employed for fine-tuning pre-trained models. Section V furnishes in-depth information regarding the datasets and the training equipment employed in our research. Section VI showcases our experimental results, addresses our research questions, and offers a performance comparison of our proposed models with state-of-the-art models. Finally, Sections VII and VIII conclude and delineate future work. 

\section{Related Works}
In this section,  we identify the most current research endeavors that employ transformer-based models for CAN attack detection. We aim to provide a comprehensive overview of the latest findings and also assess the inherent limitations of these relevant studies.

Nwafor et al. proposed a language-based intrusion detection model utilizing BERT [23]. They first trained the BERT model to comprehend the semantics within CAN messages. Subsequently, they fine-tuned the model for CAN message classification. The training procedure entailed 64\% of the data, while 20\% of the data was allocated for validation, and the remaining 16\% was dedicated to testing. Their model achieved close to 100\% accuracy, precision, Recall, and F1 score. However, the specific details of their performance were not reported.

In a related study, Natasha et al. introduced the “CAN-BERT” model [24]. Their primary objective was to design an anomaly detection model and frame the CAN message classification as a binary classification problem, differentiating between normal and abnormal messages. They adopted the BERT model and trained it using standard CAN messages, incorporating Masked Language Model (MLM) techniques. Subsequently, the model predicted the probability distribution for each CAN ID in randomly masked testing sequences. Their approach classified messages as abnormal if they lacked any CAN ID associated with a normal message. Their model achieved F1-scores ranging from 0.81 to 0.99 for different types of attacks. However, their approach heavily relied on the probability distribution of normal message CAN IDs, rendering it incapable of detecting injected message attacks that employ similar CAN IDs as normal messages. Furthermore, their model exhibited an F1 score below 0.9 for fuzzy and malfunction attacks. Another limitation of their approach is its binary classification nature, necessitating the development of an additional anomaly classification model.

Ehsan et al. introduced SecureBERT in their study [10], presenting a cybersecurity language model designed to capture text connotations within cybersecurity-related texts, such as Cyber Threat Intelligence (CTI). SecureBERT is constructed upon a pre-trained Roberta model, featuring a custom tokenizer tailored for cybersecurity domain tokens. This model was trained on an extensive corpus of cybersecurity text and was assessed using standard MLM methods. SecureBERT serves as a valuable pre-trained model imbued with domain-specific knowledge in the cybersecurity field, which we utilize in our research to develop CAN-SecureBERT.

In a separate effort,  Hugo and Louis et al. from Meta GENAI unveiled L{\scriptsize LAMA} 2 [11], a novel family of pre-trained and fine-tuned models with scales ranging from 7 billion to 70 billion parameters. The models were constructed based on the conventional transformer model as introduced in [5]. L{\scriptsize LAMA} 2 models incorporate pre-normalization using RMSNorm, employ the SwiGLU activation function, implement rotary positional embeddings, and feature grouped query attention. These models were trained using preprocessed data amounting to 2000 billion tokens, thereby encapsulating a substantial wealth of knowledge in comparison to SecureBERT. In our study, we will use L{\scriptsize LAMA} 2 to develop CAN-L{\scriptsize LAMA}2.

Consequently, as indicated by the above literature review, there is an absence of prior research endeavors that have previously employed SecureBERT and L{\scriptsize LAMA} 2 for the purpose of CAN intrusion detection and classification. Therefore, our research aims to fill a significant gap in this domain.

\section{Model Architecture}
In this section, we present a comprehensive overview of transformer architecture and the model architectures of our proposed CAN-C-BERT,  CAN-SecureBERT and CAN-L{\scriptsize LAMA}2 models.

\subsection{Transformer}
The Transformer deep learning architecture, first proposed by Google in 2017, has served as the base model for numerous NLP LLMs, including BERT, GPT, and L{\scriptsize LAMA} 2. Figure 2 illustrates the model architecture, as presented in [5], comprising an encoding and a decoding phases. The transformation process comprises 6-stack encoders and decoders,  respectively. It is akin to \textquotedblleft Transformers" components disassemble and subsequent reassemble process.  Each stack process was subdivided into eight blocks based on different initial values.  Each block possesses a \textquotedblleft transformation manual",  recording component weights and mutual relationships—a manifestation of the self-attention mechanism.

Each encoder integrates two fundamental modules: self-attention and feed-forward networks.  Each self-attention mechanism has multi-head attention heads. First, the input embeddings are generated using tokenizers, while positional information for each token is incorporated through positional encoding. The output is subsequently sent to a multi-head self-attention layer, featuring 8 multiplications of three pre-trained vectors-Key (K), Query (Q), and Value (V), to determine component weights and interrelations. The resulting weighted sum produces an attention output vector.  Afterwards, the output is processed through a fully connected feed-forward network.  In addition,  a residual connection, coupled with layer normalization, is applied to both the multi-head self-attention layer and the fully connected feed-forward layer.

The decoder has similar configurations but exhibits some distinctions from the encoder. It introduces a masked multi-head attention mechanism over the output derived from the encoder stack. Moreover, the multi-head self-attention mechanism layer within the decoder is modified to incorporate positional masking after the previous or current position, preventing attention to subsequent predictions in the sequence.  This masked adaptation is essential in autoregressive tasks (predict one token at a time),  to avoid the future positions information to influence the current prediction, which would otherwise result in data leakage.  The self-attention mechanism with multiplications of Ks and Vs from encoder states, together with Qs from decoder states, is subsequently followed by normalization through softmax function.  The softmax layer identifies the word with the highest probability as the final output.

The training methodology of the Transformer network adheres to the gradient descent algorithm, leveraging backpropagation to adjust model parameter weights by minimizing the error between predictions and actual values, thus achieving optimal learning outcomes.

\begin{figure}[!t]
\centering
\includegraphics[scale=0.5]{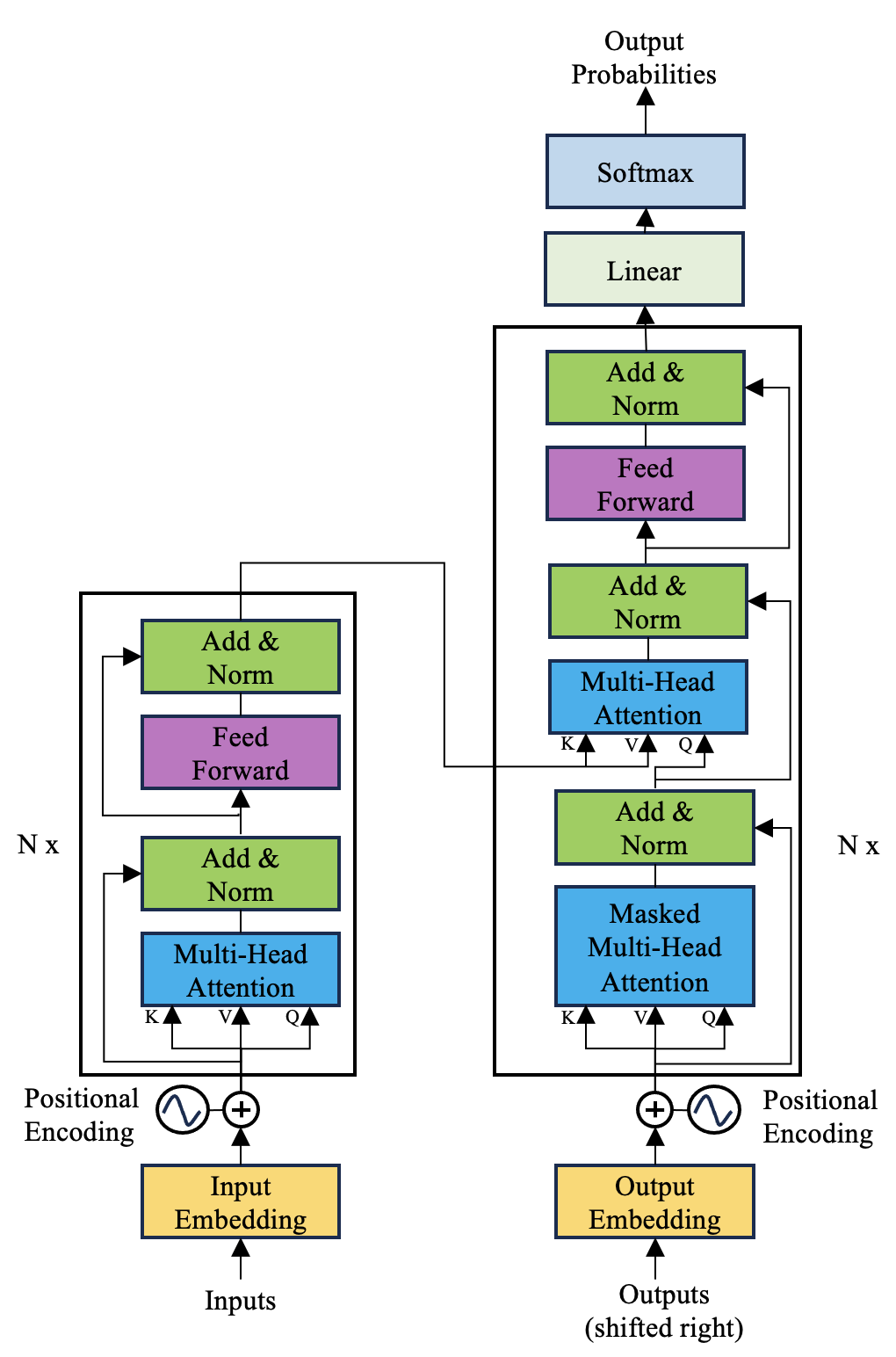}
\caption{Transformer Model Architecture [5]}
\label{Fig2}
\end{figure}

\subsection{CAN-C-BERT}

BERT, which stands for Bidirectional Encoder Representations from Transformers, is an \textquotedblleft Encoder-only" transformer introduced by Google in 2018. It undergoes training on large-scale datasets, and its robust pre-training enables fine-tuning for various downstream tasks, highlighting its versatility and effectiveness in natural language understanding.

BERT models are designed to generate deep bidirectional representations from unlabeled text by considering both left and right context across all layers. After pre-training, a BERT model can be fine-tuned for various downstream tasks, such as sequence classification, by adding a single additional output layer. In this study, we employ BERT base version model,  which has been pre-trained on the BookCorpus dataset [25]. This dataset comprises 11,038 unpublished books and English Wikipedia.  The BERT pretraining process is shown in Figure 3.

The model architecture of CAN-C-BERT model are depicted in Figures 4.  We name this fine-tuned intrusion classification model \textquotedblleft CAN-C-BERT". The \textquotedblleft C" in the middle of \textquotedblleft CAN" and \textquotedblleft BERT" is short for \textquotedblleft Classification". The architecture of the CAN-C-BERT model is particularly emphasized within the red Fine-tuning box. It incorporates a pre-trained BERT base version model and a classification head.  The pre-trained BERT base version model has 12 layers of transformer encoder blocks totaling 110 million model parameters.  The classification head is implemented as a fully connected neural network.  

\begin{figure*}[!t]
\centering
\includegraphics[scale=0.55]{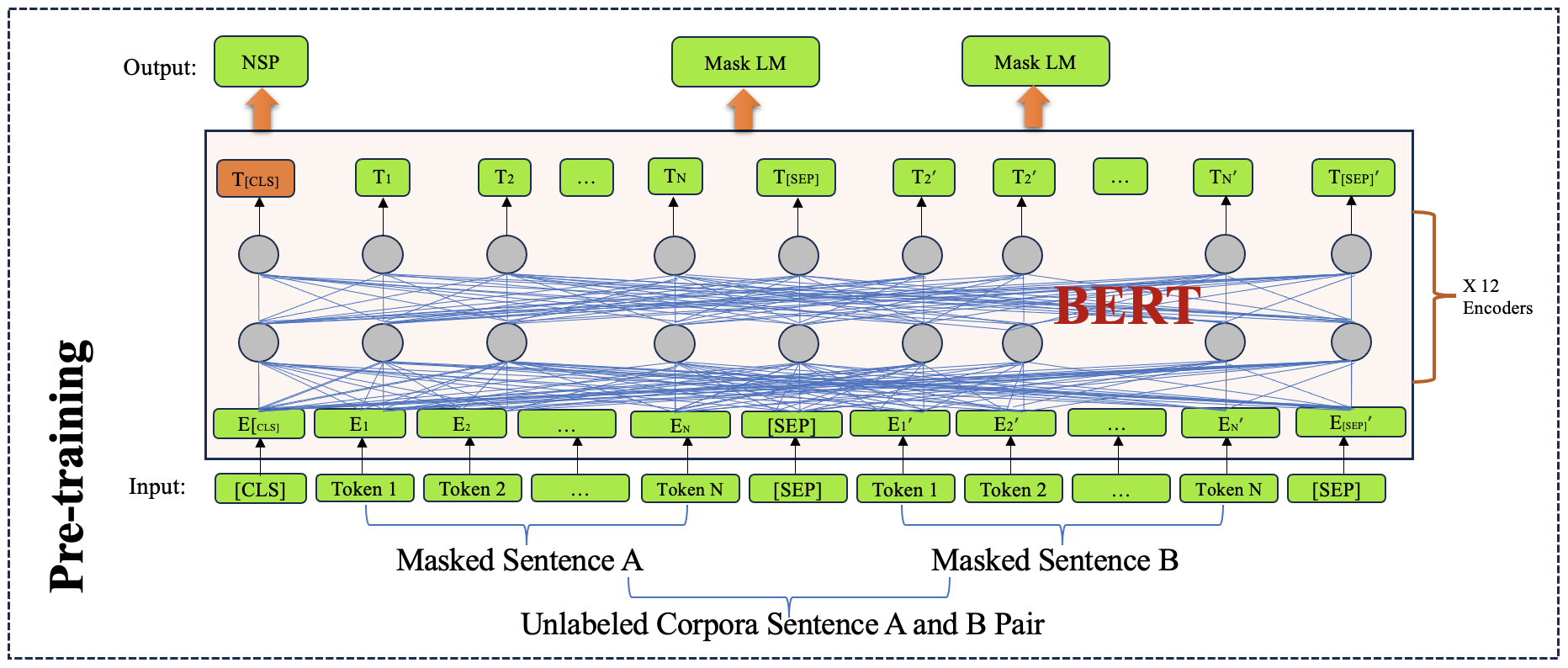}
\caption{BERT Pre-training Model [9]}
\label{Fig3}
\end{figure*}

\begin{figure*}[!t]
\centering
\includegraphics[scale=0.38]{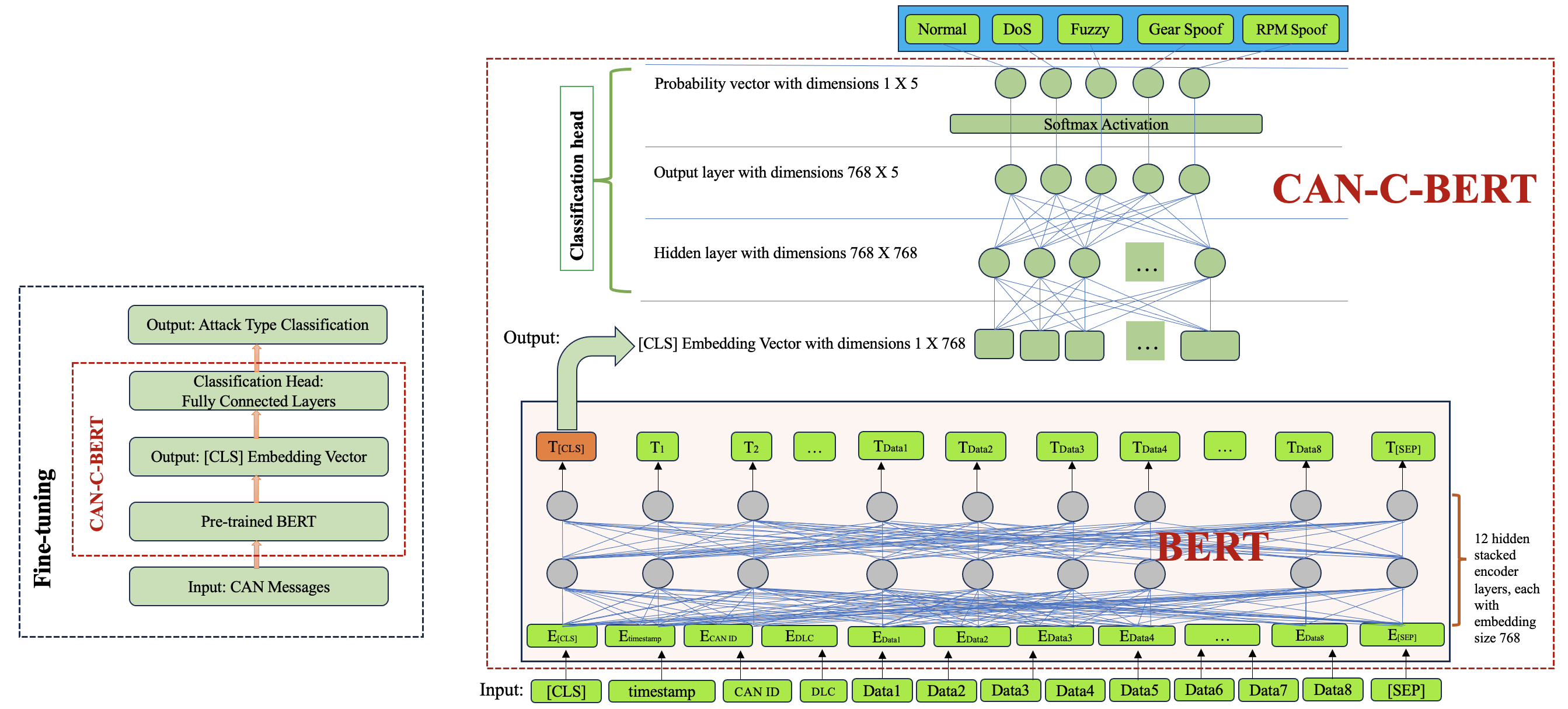}
\caption{CAN-C-BERT Fine-tuning Model}
\label{Fig4}
\end{figure*}

The first step to obtain CAN-C-BERT model is the pre-training process. The pre-training process establishes the foundational understanding of grammar and context, known as language representation. This process is characterized by its key feature—the utilization of unsupervised corpora, including consecutive sentences. Similar to the CAN-C-BERT Embedding input structure depicted in Figure 5, positional encoding information is added after segment embedding and tokenization embedding. BERT employs two primary training strategies: MLM and Next Sentence Prediction (NSP).  In MLM, each sentence masks some parts of its words, and predicting the masked words based on the context of bidirectionality, inspired by human cloze tests. After transformation, the embedding output is obtained, and the output \textquotedblleft T{\scriptsize N}" represents the prediction of the masked word. In NSP, the original input is modified to include a special token ([CLS]) at the beginning, with sentence separation indicated by the [SEP] token.

\begin{figure*}[!t]
\centering
\includegraphics[scale=0.6]{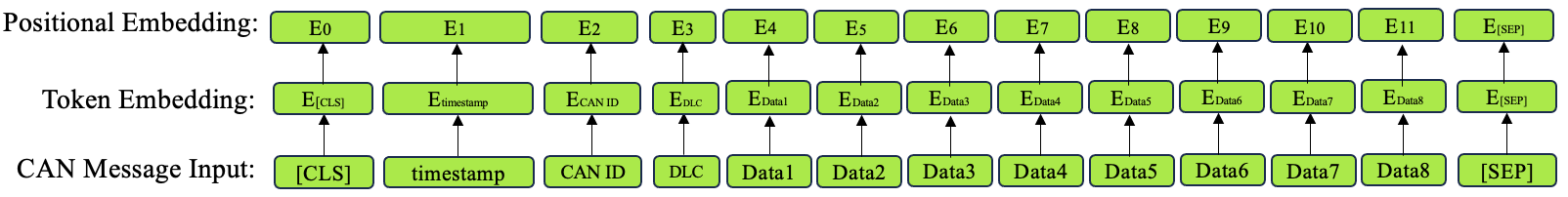}
\caption{CAN-C-BERT Embedding Input Structure: The BERT model combines Tokenization Embedding (1, n, 768) and Position Embedding (1, n, 768) through summation to obtain an input Embedding (1, n, 768) for the model.}
\label{Fig5}
\end{figure*}

The second step involved is the fine-tuning process. The fine-tuning process involves additional training tailored to specific tasks. A BERT base model consists of 12 layers of transformer encoder blocks, 768 hidden layers, and 12 self-attention heads. To train a BERT-based model for sequence classification tasks, a classification head is integrated into the embedding vector of the classification task token [CLS] derived from the BERT model. The [CLS] token summarize the sentence and represents the output of the last layer for subsequent classification. The classification head shown in Figure 4 comprises fully connected neural networks, featuring a hidden layer, an output layer, and a softmax activation layer. The [CLS] embedding vector passes through the hidden layer and connects to the subsequent output layer. The output from the output layer is then processed through a softmax activation layer to generate the probability vector. The class with the highest probability is subsequently considered the final predicted output.

For CAN attack detection and classification, the process initiates with the tokenization of the CAN messages. These tokens are then input into the pre-trained BERT model, and the output embedding for the [CLS] token becomes the input for the classification head. During training, the cross-entropy loss is computed by comparing the predicted labels to the actual ground truth labels.

\subsection{CAN-SecureBERT}
The process using CAN-SecureBERT to classify CAN messages mirrors the approach employed by CAN-C-BERT. It consists of a pre-trained SecureBERT model and a classification head, which is embodied by a fully connected neural network.  SecureBERT leverages the architecture of a pre-trained RoBERTa-base model,  featuring 12 hidden transformer and attention layers, in addition to one input layer.  This adaptation involves fine-tuning the RoBERTa-base model utilizing a substantial dataset of 98,411 cybersecurity-related textual elements (equivalent to 1 billion tokens). The model integrates a customized tokenizer based on the original RoBERTa tokenizer, effectively expanding the overall vocabulary to 50,265. This tailored tokenizer enhances the model's proficiency in extracting cybersecurity-related tokens from textual corpora. Like its name, SecureBERT compounds Security and BERT. The model's efficacy is further augmented by introducing noise into the token weights of the vocabulary during the training phase.

The model architecture of CAN-SecureBERT is represented in Figure 6, akin to CAN-C-BERT.  It incorporates a pre-trained SecureBERT model and a classification head.  The pre-trained SecureBERT model has 12 transformer blocks totaling 123 million model parameters.  The classification head is implemented as a fully connected neural network.  It is incorporated after the embedding vector of the classification token [CLS] is derived from the SecureBERT model.  It comprises a fully connected neural networks including a hidden layer, an output layer, and a softmax activation layer. The [CLS] embedding vector passes through the hidden layer and then connects to the output layer. The output from the output layer is passed into a softmax activation layer to obtain the probability vector. The class with the highest probability is the final predicted output.
\begin{figure*}[!t]
\centering
\includegraphics[scale=0.42]{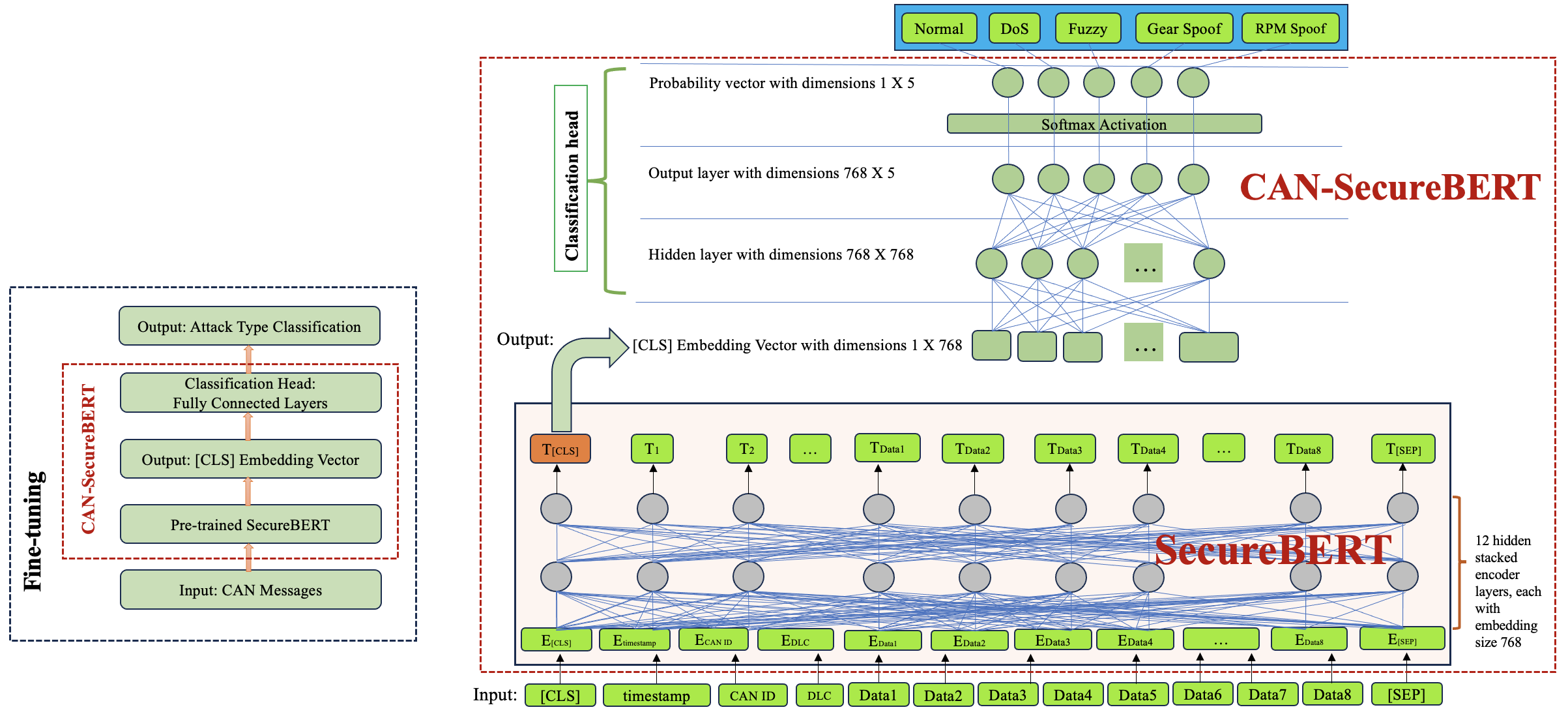}
\caption{CAN-SecureBERT Fine-tuning Model}
\label{Fig6}
\end{figure*}
\subsection{CAN-L{\scriptsize LAMA}2}
L{\scriptsize LAMA} 2, the second iteration of the L{\scriptsize LAMA} series, represents a collection of generative text models that are pre-trained and fine-tuned. It was officially introduced on July 18, 2023, as a collaborative initiative between Meta and Microsoft. The L{\scriptsize LAMA} 2 models, developed and launched by Meta, are available in three distinct sizes with parameter counts of 7 billion, 13 billion, and 70 billion. These models were trained on an extensive dataset comprising 2 trillion tokens sourced from various outlets such as web pages (CommonCrawl), open-source repository code (GitHub), Wikipedia content (in 20 different languages), public domain books, Latex source code (from scientific papers on ArXiv), and questions and answers from Stack Exchange. The dataset curation involved meticulous removal of websites containing personal data and the up-sampling of samples from reliable sources, as detailed in [11].

The model architecture of the CAN-L{\scriptsize LAMA}2 model is depicted in Figure 7. CAN-L{\scriptsize LAMA}2 model consists of a pre-trained L{\scriptsize LAMA} 2 model and the classification head. The pre-trained L{\scriptsize LAMA} 2 model utilized in this study is a half-precision model. It has 32 transformer decoder blocks totaling 7 billion model parameters.  The classification head is implemented as a fully connected neural network.  

For training the pre-trained L{\scriptsize LAMA} 2 model for CAN attack detection and classification tasks, a classification head is integrated into the embedding vector of the last token ([EOS]) (End of the Sentence)derived from the L{\scriptsize LAMA} 2 model. This classification head includes a fully connected neural network with a hidden layer, an output layer, and a softmax activation layer. The [EOS] embedding vector traverses the hidden layer and connects to the output layer. The output from the output layer undergoes a softmax activation to generate the probability vector. The final predicted output corresponds to attack type with the highest probability. For CAN attack detection and classification,  the initial step involves tokenizing the CAN messages using the L{\scriptsize LAMA} 2 tokenizer. The last token embedding will be extracted by pre-trained L{\scriptsize LAMA} 2 model. The resulting output embedding of the last token is then forwarded to the classification head. During the training process, the cross-entropy loss is calculated by comparing the predicted labels with the actual ground truth labels.

\begin{figure*}[!t]
\centering
\includegraphics[scale=0.44]{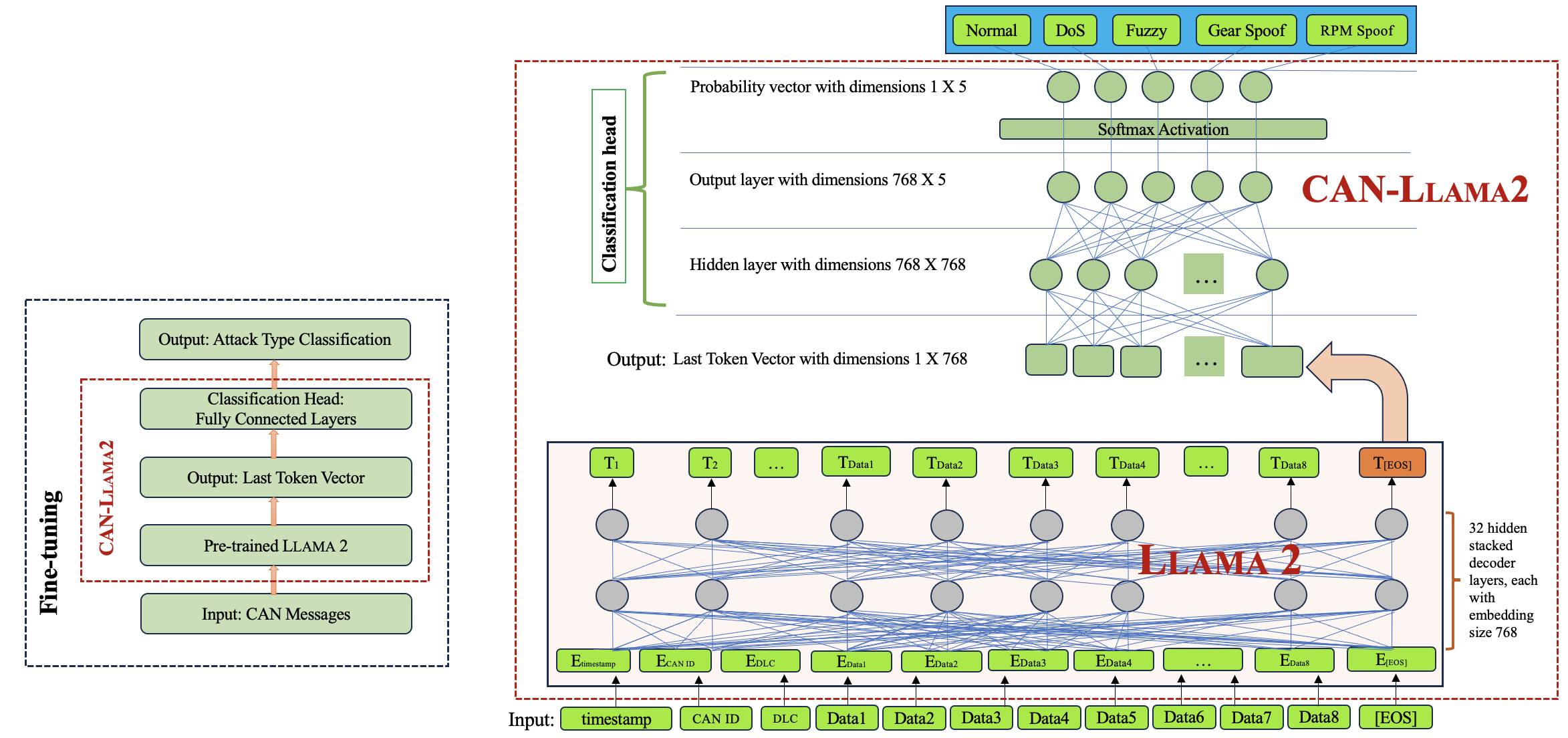}
\caption{CAN-L{\scriptsize LAMA}2 Fine-tuning Model}
\label{Fig7}
\end{figure*}

\section{Fine-tuning Process}

In this section, we explain the fine-tuning process with the aid of mathematical expressions. The main steps comprise the acquisition of class tokens from the pre-trained model,  the application of the classification head to acquire the class probability vector, the computation of the loss function, and the weight optimization procedure through the utilization of the AdamW algorithm. Additionally, we introduce methodologies that enhance the efficiency of parameter-based fine-tuning, facilitating the fine-tuning of the L{\scriptsize LAMA}2 model.

\subsection{Fine-tuning CAN-C-BERT and CAN-SecureBERT}
Initially, the individual raw CAN message is tokenized. This process involves dividing the input sequence into subword tokens using the tokenizer associated with the transformer model. Let \({X}\) symbolize the tokenized input sequence. Further, let \({model}\) denote the transformer model, where \({\theta}\) corresponds to the model's parameters. The output, denoted as  \({	Z}\) can be formulated as shown in Equation (1).

\begin{equation}
Z = \textit{model}(X, \theta)\label{eq:ft1}
\end{equation}

Atop the transformer model, a classification head is incorporated to carry out sequence classification. This classification head is composed of a hidden layer,  an output layer and subsequently followed by a softmax activation function. Let \({W}\) denote the weight matrix of the linear layer, \({b}\) signify the bias vector, and \({C}\) represent the total number of classes. The calculated predicted class probabilities \({P}_{j}\) for each class \({j}\) are calculated as expressed in Equation (2).

\begin{equation}
P_j = \frac{e^{(W_j \cdot Z + b_j)}}{\sum_{k=1}^C e^{(W_k \cdot Z + b_k)}}
\label{eq:ft2}
\end{equation}

In this study, the chosen loss function is cross-entropy loss, utilized to measure the difference between the predicted probabilities and the actual ground truth labels. Let \({{y}_{ij}}\) represent an indicator variable, which takes the value of 1 if the true label for CAN message \({i}\) corresponds to class \({j}\), and 0 otherwise. The cross-entropy loss for an individual CAN message, denoted as \({ L}_{i}\), can be defined as depicted in Equation (3).

\begin{equation}
L_i = -\sum_{j=1}^C y_{ij} \log(P_j)
\label{eq:ft3}
\end{equation}

The aggregate loss, denoted as  \({L}\), for a mini-batch of size \({N}\) is computed as the average of the individual losses \({ L}_{i}\). It can be represented as shown in Equation (4).

\begin{equation}
L = \frac{1}{N} \sum_{i=1}^{N} L_i
\label{eq:ft4}
\end{equation}

To mitigate the risk of overfitting, the optimization algorithm employed in this study is AdamW, as presented in [26]. AdamW is an adaptation of the Adam optimizer with the inclusion of weight decay regularization. It facilitates the adjustment of the model's parameters \({\theta}\) to minimize the loss function \({L}\). The weight update process can be outlined through Equation (5), where \({{\theta}_{t}}\) signifies the model parameters at iteration \({t}\), \({\eta}\) denotes the learning rate, \({{\hat{m}}_{t}}\) and \({{\hat{v}}_{t}}\) stand for the bias-corrected moving averages of the gradient and its square, \({\epsilon}\) represents a small constant introduced for numerical stability, and \({\lambda}\) denotes the weight decay coefficient.

\begin{equation}
\theta_{t+1} = \theta_t - \eta \cdot \frac{\hat{m}_t}{\sqrt{\hat{v}_t} + \epsilon} - \eta \cdot \lambda \cdot \theta_t
\label{eq:ft5}
\end{equation}

\subsection{Fine-tuning CAN-L{\scriptsize LAMA}2}

L{\scriptsize LAMA} 2, with its 7 billion parameters and half-precision, requires 14GB of GPU RAM for its operation. To facilitate training and inference while working with constrained computational resources, the study applies a technique known as LoRA from the parameter-efficient fine-tuning library [27] to fine-tune CAN-L{\scriptsize LAMA} 2.

LoRA involves the approximation of the model's weight matrices through lower-rank matrices. This approximation significantly reduces the number of parameters that need to be retrained from a pre-trained model. Let \({W}\) denote the weight matrix of a specific layer within the pre-trained model. It can be represented as \({W} \in {R}^{m*n}\), where \({m}\) signifies the number of output neurons in the layer, and \({n}\) represents the number of input neurons. The weight matrix \({W}\) can be approximated by utilizing two lower-rank matrices, \({U}\) and \({V}\), as shown in Equation (6). In this equation, \({U} \in {R}^{m*r}\) and \({V} \in {R}^{r*n}\), with \({r}\)  being the desired rank. The selection of the hyperparameter \({r}\) is guided by the balance between accuracy and the degree of compression.

\begin{equation}
W = UV
\label{eq:ft6}
\end{equation}

The primary goal is to modify the parameters of the low-rank matrices \({U}\) and \({V}\) to minimize the cross-entropy loss, as demonstrated in Equation (3), where \({L}\) denotes the task-specific loss function. In this context, \(\theta_{U}\) and \(\theta_{V}\) signify the parameters corresponding to \({U}\) and \({V}\), while \({f}\) represents the forward pass of the model, incorporating the low-rank approximation.

\begin{equation}
L_{{fine-tune}} =  L(\theta_U, \theta_V) = L(f(UV, X), Y)
\label{eq:ft7}
\end{equation}

Throughout the process of updating the model, the parameters \(\theta_{U}\) and \(\theta_{V}\) are adjusted using the AdamW optimization algorithm. This update is conducted to minimize the fine-tuning loss denoted as \({{L}_{fine-tune}}\).

\section{Performance Metrics}
In this section, we present the metrics selected for performance benchmark. We select metrics that can handle imbalanced dataset and have a direct impact to end users.

The performance evaluation of the proposed CAN-C-BERT, CAN-SecureBERT, and CAN-L{\scriptsize LAMA}2 models relies on a set of key metrics, including BA, PREC, DR or Recall, FAR, F1 score, and Model Parameter Size. The mathematical expressions for these metrics can be derived from [28], as represented in Equations (8)-(11), where \({TP}\)  stands for True Positive, \({TN}\)  signifies True Negative, \({FP}\)  represents False Positive, and \({FN}\)  denotes False Negative.

Given the inherent characteristics of the hacking dataset, it exhibits a significant imbalance, featuring notably fewer instances of attack data in comparison to normal data. As a result, the conventional accuracy metric can be misleading in such a scenario. To address this, \({BA}\)  is employed, which represents the mean accuracy of individual class predictions. \({BA}\)  offers a more reliable measure of model performance, particularly suited for datasets characterized by an imbalanced class distribution.

\begin{equation}
BA = \frac{1}{C} \sum_{i=1}^{C} \frac{TP_i}{TP_i + FN_i}
\label{eq:ft8}
\end{equation}

The \({PREC}\)  measures the accuracy of positive predictions generated by a model.
\begin{equation}
PREC = \frac{{TP}}{{TP + FP}}
\label{eq:ft9}
\end{equation}

\({DR}\) , commonly known as Recall,  assesses the proportion of \({TP}\)  predictions relative to the entirety of actual positive cases.
\begin{equation}
DR = \frac{TP}{TP + FN}
\label{eq:ft10}
\end{equation}

\({FAR}\)  quantifies the proportion of \({FP}\)  predictions relative to the entire set of actual negative cases. This metric holds significant importance, particularly in the context of an IDS. For instance, a model with a \({FAR}\)  of 1e-3 may be deemed exceptional performance in the majority of ML tasks. However, in the case of IDS, the practical implications become apparent. Consider the scenario where millions of CAN messages are generated daily by a multitude of vehicles. To illustrate, let us assume a daily volume of 10 million messages. With a \({FAR}\)  of 1e-3, the number of false alarm messages would amount to 10,000. Managing and triaging such a volume on a daily basis becomes operationally impractical for a VSOC team.

\begin{equation}
FAR = \frac{FP}{FP + TN}
\label{eq:ft11}
\end{equation}

The F1 score represents the harmonic mean of \({PREC}\)  and \({DR}\)  or Recall. This metric furnishes a balanced and comprehensive measure of a model's performance.
\begin{equation}
F1 = \frac{2 \cdot PREC \cdot DR}{PREC + DR}
\label{eq:ft12}
\end{equation}

\section{Datasets}

In this study, we employed the Car Hacking Dataset, which was gathered from Hyundai's YF Sonata and made available by Song et al. [13]. This dataset encompasses three distinct categories of attacks, namely, DoS attacks, Fuzzy attacks, and Spoofing attacks. These datasets were curated by capturing CAN traffic through the On-Board Diagnostics II (OBD-II) port of a real vehicle while simulated message attacks were actively being executed. The data attributes within the dataset encompass the following fields: Timestamp, CAN ID, Data Length Code (DLC), DATA[0], DATA[1], DATA[2], DATA[3], DATA[4], DATA[5], DATA[6], DATA[7], and Flag. Further insights into each attack type, as detailed in [13], are outlined as follows:

\begin{enumerate}
\item \textit{DoS Attack}: The DoS attack involves the injection of high-priority CAN messages, exemplified by the `0x000' CAN ID packet, at brief intervals. Specifically, these `0x000' CAN ID messages were injected every 0.3 milliseconds.
\item \textit{Fuzzy Attack}: The Fuzzy attack entails the injection of messages featuring spoofed, randomly generated CAN ID and DATA values. This injection was carried out at intervals of 0.5 milliseconds, including messages with randomized CAN ID and data values.
\item \textit{RPM/Gear Attack}: The RPM/Gear attack revolves around the injection of messages associated with specific CAN IDs pertinent to RPM and Gear information. Messages related to RPM and Gear were introduced at intervals of 1 millisecond.
\end{enumerate}

These attacks were executed by introducing messages into the CAN network, aiming to replicate real-world scenarios and behaviors within the Hyundai YF Sonata. An overview of the Car Hacking dataset is presented in Table I.

\begin{table}[!t]
   \caption{CAR HACKING DATASET \label{tab:table1}}
   \centering
   \begin{tabular}{|c||c|c|c|} 
   \hline
     \textbf{Attack Type} & \textbf{Messages} & \textbf{Normal} & \textbf{Injected}\\
     \hline
     DoS Attack & 3665771 & 3078250 & 587521\\
     Fuzzy Attack & 3838860 & 3347013 & 491847\\
     Spoofing  Drive Gear & 4443142 & 3845890 & 597252\\
     Spoofing  RPM Gauze & 4621702 & 3966805 & 654897\\
     Attack-Free (Normal) & 988987  & 988872 & NA\\
      \hline
   \end{tabular}
\end{table}

\section{Experiments and Results}
In this section, we present how to process the dataset to train and evaluate our proposed models. We show the experimental setup, hyperparameters employed, as well as model complexity.  Subsequently, we delve into a comprehensive discussion of the experimental results and emphasize noteworthy observations.
\subsection{Fine-tuning Hyperparameters}
We partitioned the Car Hacking Dataset into a 70\% training dataset and a 30\% test dataset. In practical applications, collecting data from vehicles is often a challenging endeavor. It is crucial to explore how the quantity of training data affects model performance. Therefore, our proposed models were trained and validated using subsets amounting to 1\% and 10\% of the training dataset, which were randomly selected from the entire dataset.  The random selection process maintained a balanced representation of data, with the normal data comprising a ratio ten times smaller than other attack-type data. For instance, when the selected attack types amounted to 1\%, the normal dataset selected was 0.1\%, ensuring a balanced distribution of normal and attack-type data. Subsequently, the selected dataset was further divided into a 70\% training portion and a 30\% validation portion. The entire 30\% test dataset from the Car Hacking dataset was reserved to evaluate the trained and validated models.

The hardware used for training in this study possesses the following specifications: AMD Ryzen 9 5900X 12-Core Processor with a clock speed of 3.70 GHz, 128GB of RAM, and two Nvidia RTX 3090 GPUs, each equipped with 24GB of VRAM and interconnected using the Nvidia SLI bridge.

For the training hyperparameters, identical settings are applied to both CAN-C-BERT and CAN-SecureBERT. These models utilize a training batch size of 4 and a validation batch size of 32. The learning rate is set at 5e-5, while the weight decay is configured at 0.01. As for CAN-L{\scriptsize LAMA}2, the training batch size is set to 4, the validation batch size is 16, and a gradient accumulation step of 4 is implemented. The learning rate for CAN-L{\scriptsize LAMA}2 is established at 3e-5, and the weight decay is fixed at 0.01. Moreover, to facilitate the training of CAN-L{\scriptsize LAMA}2 while accommodating limited computational resources, the model parameters are loaded in 4-bit precision. LoRA, as discussed in Section IV, is incorporated into the training pipeline. Specifically, the LoRA attention dimension is set to 16, the alpha parameter for LoRA scaling is established at 64, a dropout probability of 0.1 is applied to LoRA layers, and the bias is maintained at a value of 0. All three models are trained for a total of 10 epochs.

\subsection{Model Complexity}
Table II presents a comparative analysis of fine-tuned model sizes and parameters. Both CAN-C-BERT and CAN-SecureBERT offer the ability to fine-tune all their parameters during training. Notably, when employing a 1 percent training dataset, the training times for these models are approximately 4 and 5 minutes, respectively. In contrast, training the CAN-L{\scriptsize LAMA}2 model requires considerably more time, with a training duration of 118 minutes. However, it is important to note that the inference speed of CAN-L{\scriptsize LAMA}2 is approximately eight times slower than the other two models. This difference is primarily attributed to computational resource limitations, given that CAN-L{\scriptsize LAMA}2 concludes 7 billion parameters. After implementing LoRA, it becomes possible to fine-tune approximately 40 million parameters from the linear layers of the model. Nevertheless, it is worth emphasizing that the computational demands of CAN-L{\scriptsize LAMA}2 in terms of matrix multiplications, activations, and other mathematical operations are significantly higher than those of the other two models. Furthermore, due to restrictions related to GPU memory size, CAN-L{\scriptsize LAMA}2 can only accommodate a maximum batch size of 16 CAN messages for validation. This limitation not only affects memory access times but also gives rise to memory bottlenecks.

Another noteworthy observation is that only 0.57\% of the parameters are altered in the CAN-L{\scriptsize LAMA}2 model during fine-tuning. This implies that the majority of the original L{\scriptsize LAMA}2 model's parameters remain unchanged. Consequently, CAN-L{\scriptsize LAMA}2 can be repurposed for other language-related tasks. The VSOC team can utilize the same model and undertake various downstream tasks by fine-tuning pre-trained models and incorporating adapter heads.

\begin{table*}[!t]
\caption{MODEL SIZE\label{tab:table2}}
\centering
   \begin{tabular}{|c||c|c|c|c|} 
         \hline
     \textbf{Model Type} & \textbf{Model Parameters} & \textbf{Trainable Parameters}& \textbf{Training Time}& \textbf{Inference Speed} \\
     \hline
     CAN-C-BERT& 110 Million & 110 Million & 4  mins & 894 messages/s \\
     CAN-SecureBERT& 123 Million & 123 Million & 5 mins & 965 messages/s  \\
     CAN-L{\scriptsize LAMA}2 & 7 Billion & 40 Million & 118 mins & 14 messages/s \\
      \hline
   \end{tabular}
\end{table*}

\subsection{Results}
In this section, we conduct a thorough analysis of our results to address our research questions and gain insights into the performance of our proposed models. The validation results, including training loss, validation loss, BA, PREC, DR, and F1 score, are visualized in Figures 8-13.

\subsubsection{Training Loss and Validation Loss}
Training loss and validation loss result are shown in Figure 8 and 9. All models have training and validation loss close to 0 within 10 epochs, which indicates that CAN-C-BERT, CAN-SecureBERT, and CAN-L{\scriptsize LAMA}2 all converge within ten epochs.  Figure 8 clearly illustrates that the models trained with 10\% of the data converge more swiftly than those trained with only 1\% of the data.  All six training losses are close to 0, indicating that they have all converged. However, for validation loss from Figure 9, the models trained with 10\% data have loss close to 0, which are much less than the models trained with 1\% data. This indicates that models trained with 10\% exhibit more accurately.  The validation loss is close to 0.  It means these models can perform effectively on the unseen data within the validation dataset.

\begin{figure}[!t]
\centering
\includegraphics[scale=0.42]{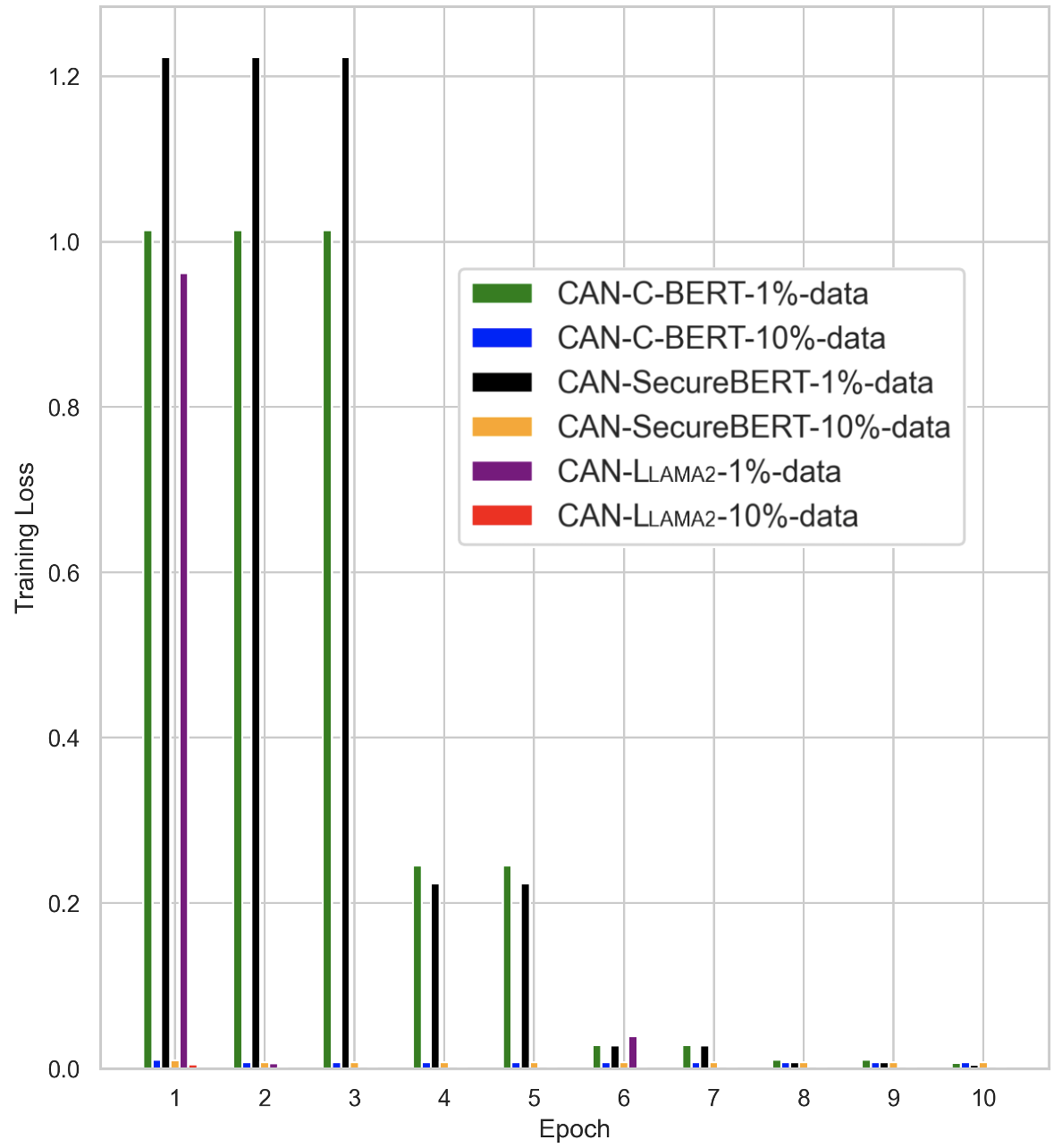}
\caption{Train Loss Comparison between CAN-C-BERT, CAN-SecureBERT, and CAN-L{\scriptsize LAMA}2 with 1\% and 10\% Training Data}
\label{Fig9}
\end{figure}

\begin{figure}[!t]
\centering
\includegraphics[scale=0.42]{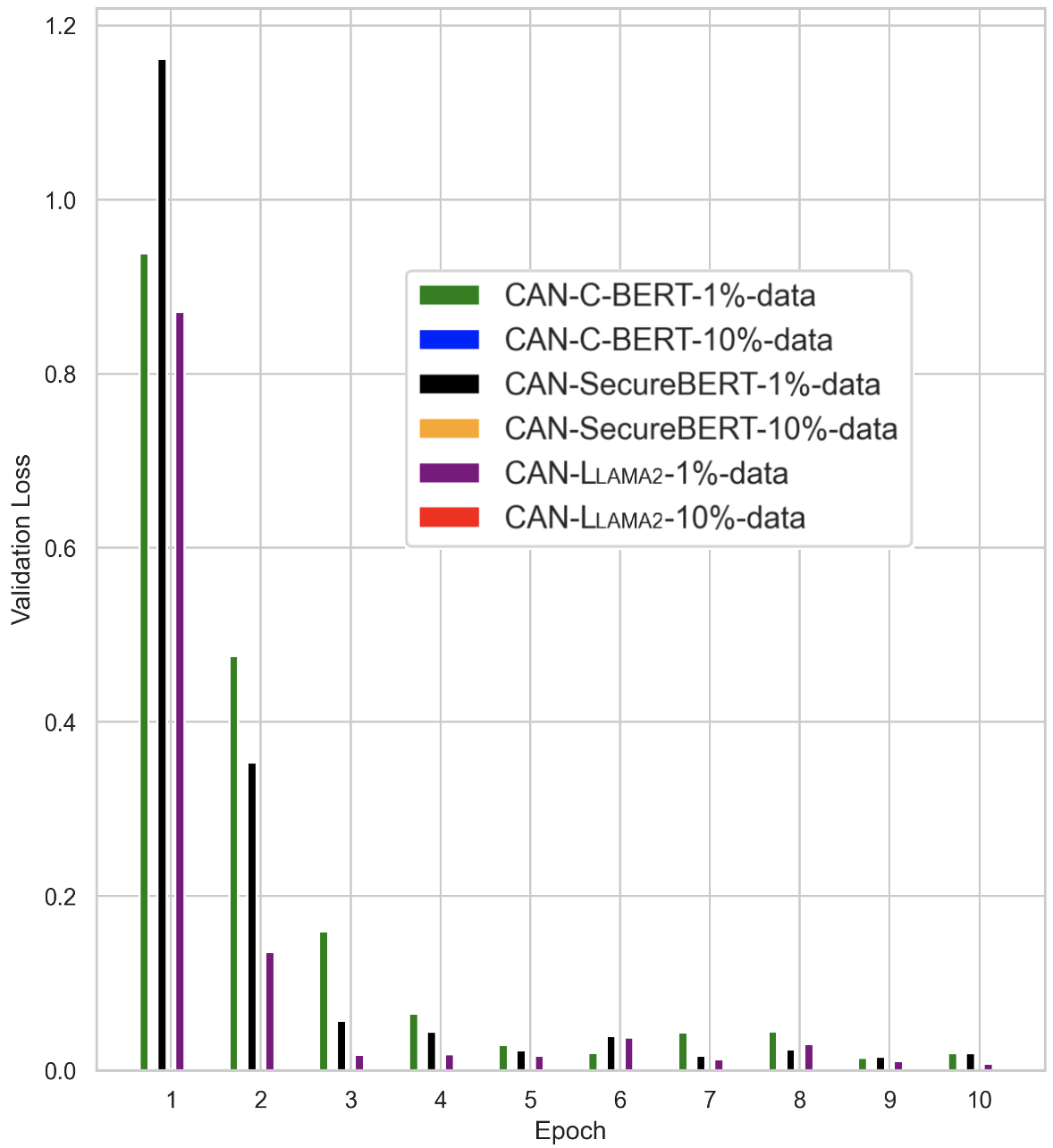}
\caption{Validation Loss Comparison between CAN-C-BERT, CAN-SecureBERT, and CAN-L{\scriptsize LAMA}2 with 1\% and 10\% Training Data}
\label{Fig10}
\end{figure}

\subsubsection{Balanced Accuracy, Precision, Detection Rate and F1 Score}
The BA, PREC, DR, and F1 score are plotted in Figure 10 to 13. All 6 models have achieved close to the value 1 result for these metrics. It indicates all models are very accurate. From the plots of model performance using 1\% training data in Figures 10-13, we can observe CAN-L{\scriptsize LAMA}2 exhibits superior early-stage performance and converges to the value of 1.0 more rapidly compared to the other two models. The models trained with 10\% data perform better than the models trained with 1\% data for all metrics. Among all six models, CAN-L{\scriptsize LAMA}2 trained with 10\% data performs the best. CAN-SecureBERT trained with 10\% data is the second best among all metrics.

\begin{figure}[!t]
\centering
\includegraphics[scale=0.33]{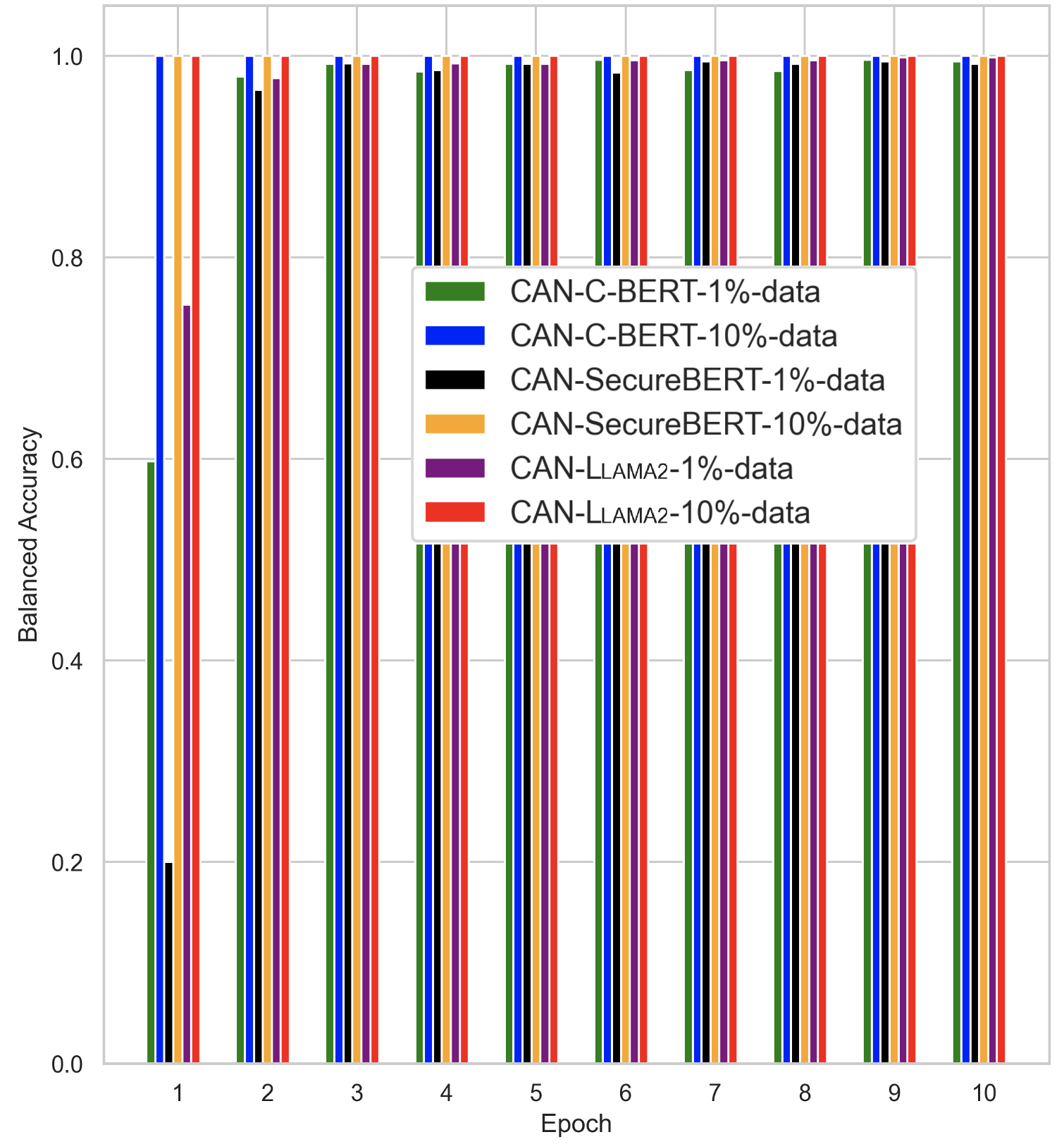}
\caption{Balanced Accuracy Comparison between CAN-C-BERT, CAN-SecureBERT, and CAN-L{\scriptsize LAMA}2 with 1\% and 10\% Training Data}
\label{Fig11}
\end{figure}

\begin{figure}[!t]
\centering
\includegraphics[scale=0.42]{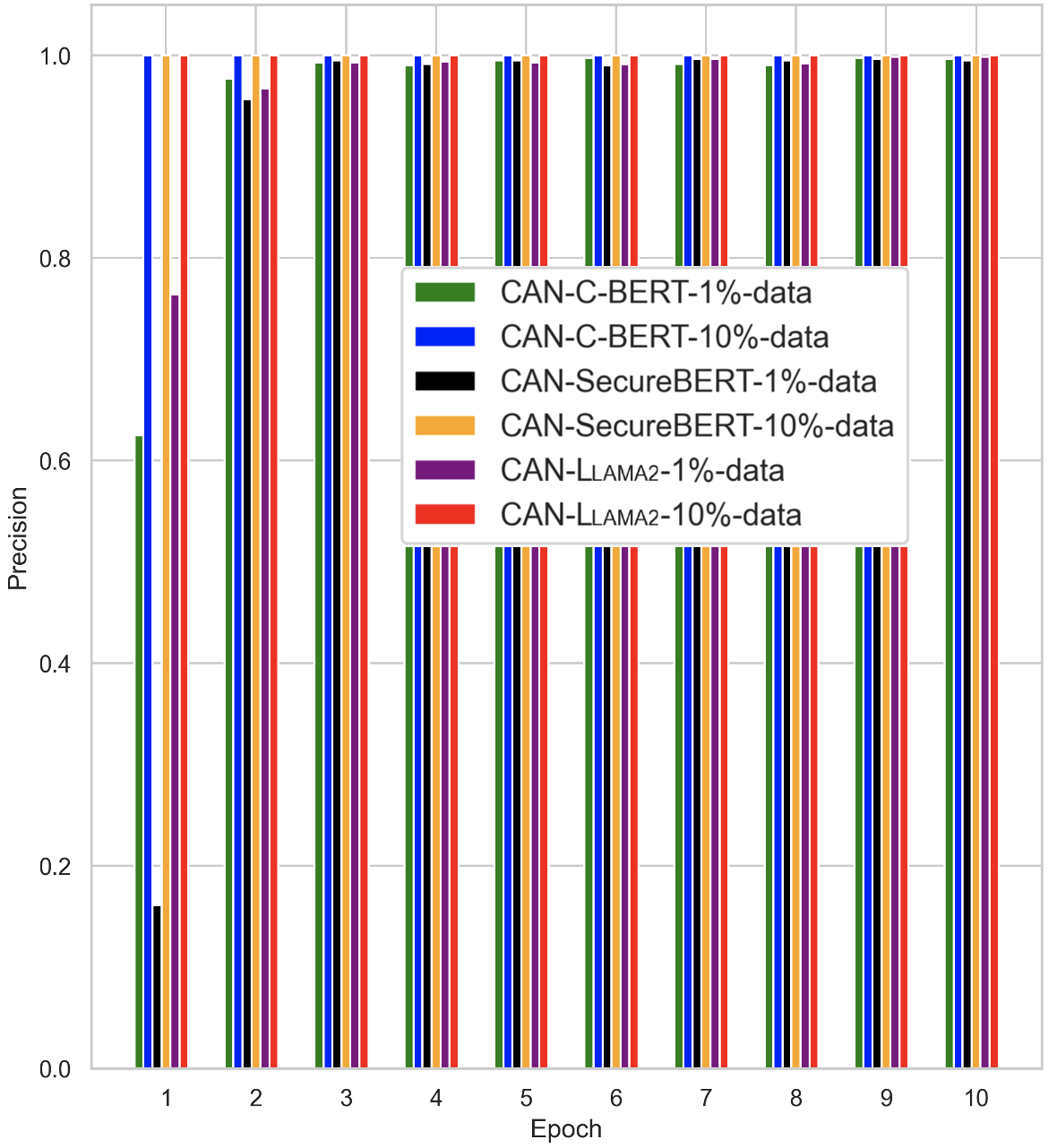}
\caption{Precision Comparison between CAN-C-BERT, CAN-SecureBERT, and CAN-L{\scriptsize LAMA}2 with 1\% and 10\% Training Data}
\label{Fig12}
\end{figure}

\begin{figure}[!t]
\centering
\includegraphics[scale=0.42]{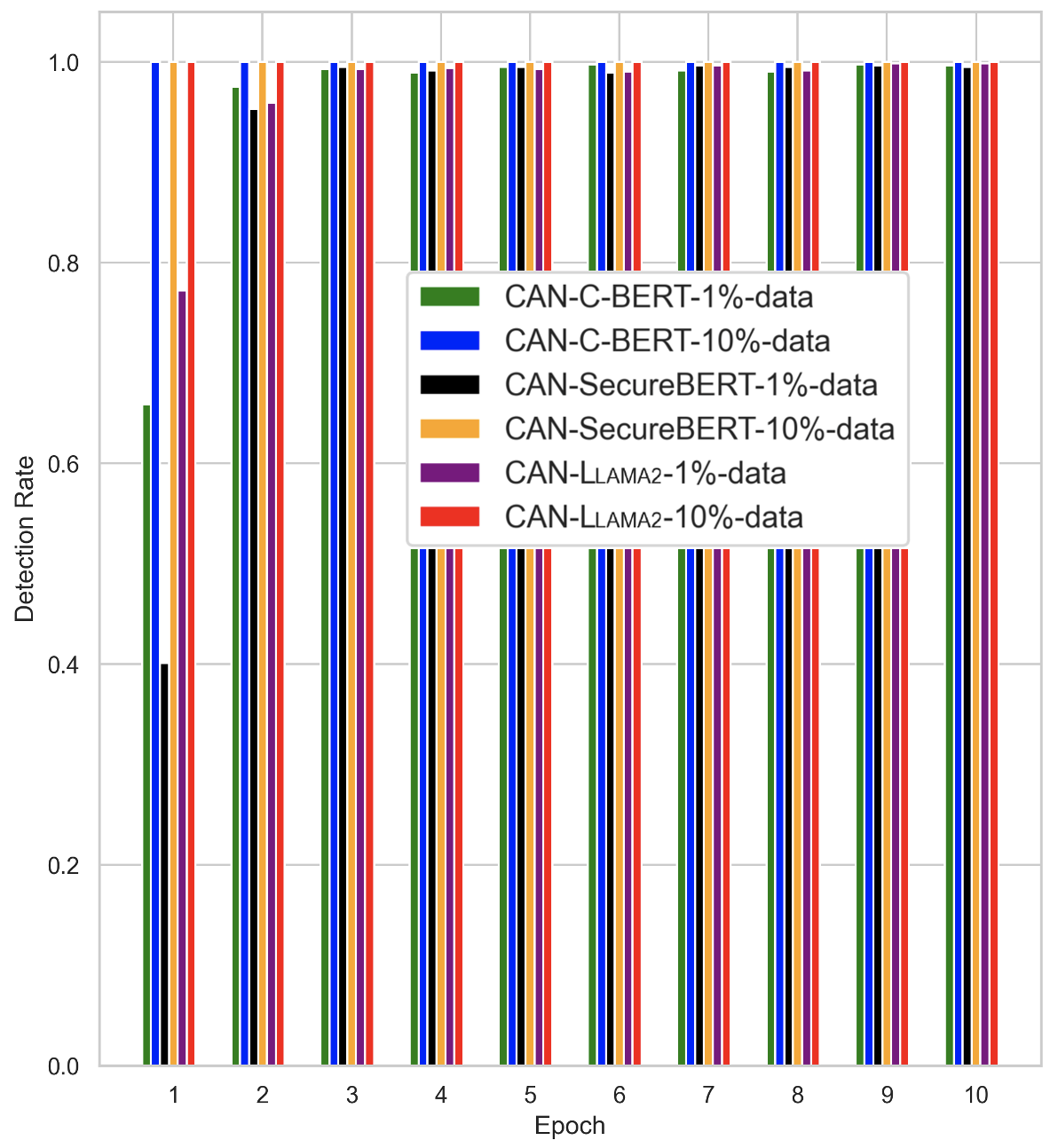}
\caption{Detection Rate Comparison between CAN-C-BERT, CAN-SecureBERT, and CAN-L{\scriptsize LAMA}2 with 1\% and 10\% Training Data}
\label{Fig13}
\end{figure}

\begin{figure}[!t]
\centering
\includegraphics[scale=0.42]{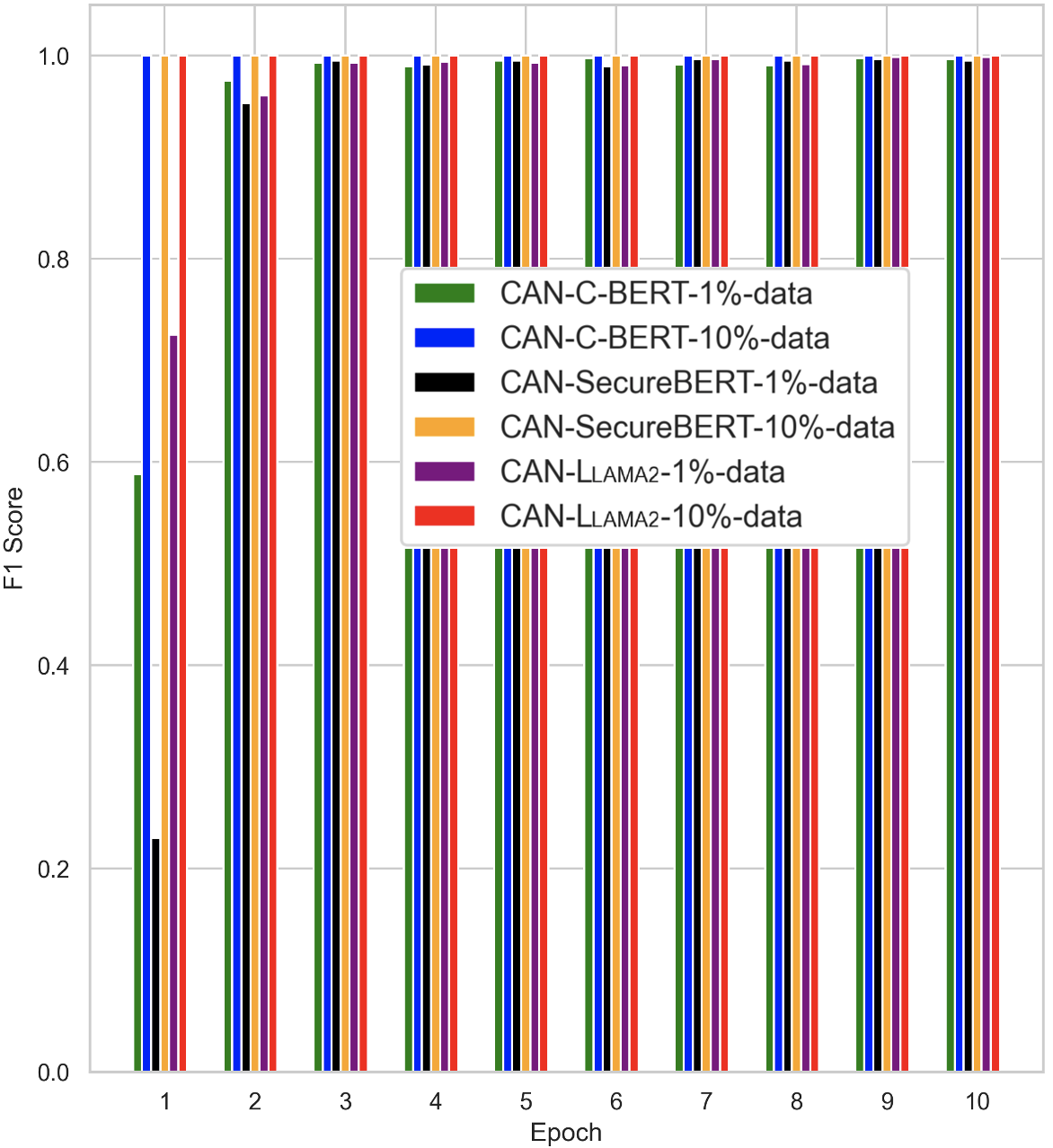}
\caption{F1 Score Comparison between CAN-C-BERT, CAN-SecureBERT, and CAN-L{\scriptsize LAMA}2 with 1\% and 10\% Training Data}
\label{Fig14}
\end{figure}

Table III provides a comprehensive summary of the models' performance after testing, with a comparison to state-of-the-art models. CAN-L{\scriptsize LAMA}2, trained with 10\% of the data, outperforms all other state-of-the-art models, achieving BA, PREC, DR, and F1 scores of 0.999993, with a FAR of 3.1e-6. This remarkable performance implies that, with a dataset of 10 million CAN messages, only 31 messages are expected to be false alarms. The second-best model is CAN-SecureBER, also trained with 10\% of the data, achieving BA, PREC, DR, and F1 scores of 0.999991. In this case, there are 35 expected false alarms for 10 million CAN messages. Both models significantly outperform  MTH-IDS and CAN-C-BERT, thereby reducing the workload for VSOC team during the triage process. These findings underscore the effectiveness of a more complex model with pre-trained knowledge for vehicle network intrusion detection, surpassing an IDS designed with a combination of multiple models and algorithms. It demonstrates that a single base model, with fine-tuned parameter sets or adapters, can handle multiple downstream tasks.

\begin{table*}[!t]
\caption{MODEL PERFORMANCE\label{tab:table3}}
\centering
\begin{tabular}{|c||c|c|c|c|c|}
      \hline
  \textbf{Model Type} & \textbf{BA} & \textbf{PREC}  & \textbf{DR} & \textbf{FAR}& \textbf{F1}\\
     \hline\     
     DCNN[13] & 0.999300 & - & 0.998400 & 1.60e-3 & 0.999100 \\
     E-MLP[14] & 0.990000 & 0.990000 & 0.990000 & 3.00e-3 & 0.990000 \\
     GIDS[15] & 0.975250 &0.976250  & 0.986500 & 1.00e-3 & 0.987900 \\
     Transformer Sequential CAN ID[16] & - & 0.998980 & 999350 & 5.80e-4 & 0.999170 \\
     KNN[17] & 0.974000 & - & 0.963000 & 5.30e-2 & 0.934000 \\
     SVM[17] & 0.965000 & - & 0.957000 & 4.80e-2 & 0.933000 \\
     XYF-K[18] & 0.991000 & - & 0.983900 & - & 0.987900 \\
     SAIDuCANT[19] & 0.872100 & - & 0.866600 & 1.70e-2 & 0.920000 \\
     SSAE[20] & -  & - & 0.985000 & 1.76e-2 & 0.920000 \\
     LSTM-Autoencoder[21] & 0.990000 & - & 0.990000 & - & 0.990000 \\
     MTH-IDS[12] & 0.999990 & - & 0.999990 & 6.00e-4 & 0.999990 \\
     CAN-C-BERT-1\%-data & 0.991377 & 0.992874 & 0.991377 & 9.90e-3 & 0.991773 \\
     CAN-SecureBERT-1\%-data& 0.999718& 0.999718 & 0.999718 & 1.30e-4 & 0.999718\\
     CAN-L{\scriptsize LAMA}2-1\%-data & 0.999587& 0.999588 & 0.999587 & 3.10e-4 & 0.999587 \\
     CAN-C-BERT-10\%-data& 0.999965 & 0.999965& 0.999965 & 3.60e-5 & 0.999965 \\
     CAN-SecureBERT-10\%-data& 0.999991& 0.999991 & 0.999991 & 3.50e-6 & 0.999991\\
     \textbf{CAN-L{\scriptsize LAMA}2-10\%-data} & \textbf{0.999993} & \textbf{0.999993} & \textbf{0.999993} & \textbf{3.10e-6} & \textbf{0.999993} \\
      \hline
\end{tabular}
\end{table*}

The detailed model performance for the classification of each attack type is listed in Tables IV, V, and VI. CAN-SecureBERT and CAN-L{\scriptsize LAMA}2 both achieve 100\% performance for DoS, gear spoofing, and RPM spoofing attacks. However, both models make some few incorrect predictions for fuzzy attacks, which involve the injection of randomly spoofed messages.

\begin{table*}[!t]
\caption{ATTACK CLASSIFICATION PERFORMANCE - CAN-C-BERT\label{tab:table4}}
\centering
   \begin{tabular}{|c||c|c|c|c|c|} 
         \hline
     \textbf{Attack Type} & \textbf{Validation Instances} & \textbf{PREC}  & \textbf{DR} & \textbf{FAR}& \textbf{F1}\\
     \hline
     DoS  & 117674 & 0.999230 & 1.0 & 3.1e-5 & 0.999614 \\
     Fuzzy & 98867 & 0.999868 & 0.999888 & 4.5e-6 & 0.999878\\
     Gear Spoofing & 119707 & 1.0 & 1.0 & 0 & 1.0 \\
     RPM Spoofing & 131324 & 1.0 &1.0 & 0 & 1.0 \\
      \hline
   \end{tabular}
\end{table*}

\begin{table*}[!t]
\caption{ATTACK CLASSIFICATION PERFORMANCE - CAN-C-SECUREBERT\label{tab:table5}}
\centering
   \begin{tabular}{|c||c|c|c|c|c|} 
         \hline
     \textbf{Attack Type} & \textbf{Validation Instances} & \textbf{PREC}  & \textbf{DR} & \textbf{FAR}& \textbf{F1}\\
     \hline
     DoS  & 117674 & 1.0 & 1.0 & 0 &1.0 \\
     Fuzzy & 98867 & 0.999898 & 0.999828 & 3.5e-6 & 0.999878\\
     Gear Spoofing & 119707 & 1.0 & 1.0 & 0 & 1.0 \\
     RPM Spoofing & 131324 & 1.0 &1.0 & 0 & 1.0 \\
      \hline
   \end{tabular}
\end{table*}

\begin{table*}[!t]
\caption{ATTACK CLASSIFICATION PERFORMANCE - CAN-L{\scriptsize LAMA}2\label{tab:table6}}
\centering
   \begin{tabular}{|c||c|c|c|c|c|} 
         \hline
     \textbf{Attack Type} & \textbf{Validation Instances} & \textbf{PREC}  & \textbf{DR} & \textbf{FAR}& \textbf{F1}\\

     \hline
     DoS  & 117674 & 1.0 & 1.0 & 0 & 1.0 \\
     Fuzzy & 98867 & 0.999909 & 0.999858 & 3.1e-6 & 0.999883\\
     Gear Spoofing & 119707 & 1.0 & 1.0 & 0 & 1.0 \\
     RPM Spoofing & 131324 & 1.0 &1.0 & 0 & 1.0 \\
      \hline
   \end{tabular}
\end{table*}

\subsection{Discussions}

Based on the above results,  there is a significant difference about the changing trends of the training loss and the validation loss from Figure 8 and 9. This difference indicates that the models trained with a larger dataset tend to generalize more effectively than those trained with a smaller dataset. This conclusion is substantiated by the analysis of other model performance metrics, including BA, PREC, DR, and F1 score, where the models trained with 10\% of the data consistently outperform their 1\% data counterparts.

All proposed models achieve exceptional performance in CAN message log classification. These models operate directly on raw, text-based CAN messages, and their BA, PREC, DR, and F1 score all exceed 0.99, as shown in Figures 10-13. This feature indicates that transformer-based models can directly classify CAN message logs without the need for feature engineering and data preprocessing.

When looking at the model performance with 1\% data from Figure 10 to 13, CAN-L{\scriptsize LAMA}2 converges faster than the rest of models with 1\% data. This  indicates that larger models, such as CAN-L{\scriptsize LAMA}2, have a better capability to capture complex CAN message patterns. The increased number of model layers in CAN-L{\scriptsize LAMA}2 facilitates more efficient information sharing across layers due to its larger parameter size, contributing to its faster convergence.

The model with more pre-trained knowledge from transformer models perform better compared to models with less pre-trained knowledge. L{\scriptsize LAMA}2 is pre-trained on a larger dataset compared to BERT and SecureBERT. To capture more pre-trained knowledge, L{\scriptsize LAMA}2 has 54 times more parameters than BERT and SecureBERT. However, only 40 million parameters are allowed to be fine-tuned after applying LoRa, with the remaining parameters being frozen. Despite the reduced number of fine-tuned parameters, CAN-L{\scriptsize LAMA}2 outperforms CAN-C-BERT and CAN-SecureBERT. This observation indicates that a larger transformer-based model, trained on an extensive dataset, can enhance CAN classification performance.

CAN-SecureBERT uses SecureBERT which is primarily trained with cybersecurity-focused data. While it exhibits slightly worse performance than CAN-L{\scriptsize LAMA}2, it outperforms CAN-C-BERT. However, SecureBERT has a parameter size that is 20\% larger than CAN-C-BERT. Consequently, it remains uncertain whether the performance difference can be attributed to the pre-trained domain knowledge or the increased number of parameters in the model.

\section{Conclusion}

In this study, we propose a novel approach for CAN intrusion detection and attack classification by fine-tuning pre-trained transformer-based models. Three distinct models, namely CAN-C-BERT, CAN-SecureBERT, and CAN-L{\scriptsize LAMA}2 have been developed. Our proposed models can directly use CAN message logs and eliminate the need to perform data preprocessing. After trained by using pre-balanced CAN dataset,  their  performances have been compared against state-of-the-art models. CAN-L{\scriptsize LAMA}2 exhibits the highest level of performance, surpassing all empirical state-of-the-art IDS systems. CAN-SecureBERT stands as the second-best model. The leading model, CAN-L{\scriptsize LAMA}2, achieves outstanding results with a BA, PREC, DR, and F1 score of 0.999993, accompanied by an impressively low FAR of 3.1e-6. This FAR is approximately 52 times better than that of MTH-IDS, clearly outperforming all other state-of-the-art models.  Overall, our study advances the field of CAN IDS, offering insights into model design, performance, and adaptability for cybersecurity applications.

\section{Future Works}

One of the key limitations of our research is the computational resource constraints. In our forthcoming research, we will explore methods to further reduce the model size and enhance inference speed of the proposed CAN-L{\scriptsize LAMA}2 model.


\begin{thebibliography}{1}
\bibliographystyle{IEEEtran}


\bibitem{ref1} \textquotedblleft UN Regulation No. 155 - CyberSecurity and CyberSecurity Management System," {\it{UNECE}}, [Online]. Available: https://unece.org/transport/documents/2021/03/standards/un-regulation-no-155-cyber-security-and-cyber-security. [Accessed: Oct. 30, 2023].

\bibitem{ref2} Wu, W., Li, R., Xie, G., An, J., Bai, Y., Zhou, J., and Li, K., \textquotedblleft A survey of intrusion detection for in-vehicle networks,"   {\it{IEEE Transactions on Intelligent Transportation Systems}}, vol. 21, no. 3, pp. 919-933, 2019.

\bibitem{ref3} K. T. Cho and K. G. Shin, \textquotedblleft  Viden: Attacker identification on in-vehicle networks," in {\it{Proceedings of the 2017 ACM SIGSAC Conference on Computer and Communications Security}}, pp. 1109-1123, 2017.

\bibitem{ref4} H. M. Song, H. R. Kim, and H. K. Kim, \textquotedblleft Intrusion detection system based on the analysis of time intervals of CAN messages for in-vehicle network," in {\it{2016 International Conference on Information Networking (ICOIN)}}, pp. 63-68, IEEE, 2016.

\bibitem{ref5} A. Vaswani, N. Shazeer, N. Parmar, J. Uszkoreit, L. Jones, A. N. Gomez,  L. Kaiser, and I. Polosukhin, \textquotedblleft Attention is all you need," in {\it{Advances in Neural Information Processing Systems 30}}, 2017.

\bibitem{ref6} N. S. Keskar, B. McCann, L. R. Varshney, C. Xiong, and R. Socher, \textquotedblleft Ctrl: A conditional transformer language model for controllable generation," arXiv preprint arXiv:1909.05858, 2019.

\bibitem{ref7} S. Niu, Y. Liu, J. Wang, and H. Song, \textquotedblleft A decade survey of transfer learning (2010–2020), " {\it{IEEE Transactions on Artificial Intelligence}}, vol. 1, no. 2, pp. 151-166, 2020.

\bibitem{ref8} C. Raffel, N. Shazeer, A. Roberts, K. Lee, S. Narang, M. Matena, Y. Zhou, W. Li, and P. J. Liu, \textquotedblleft Exploring the limits of transfer learning with a unified text-to-text transformer," {\it{The Journal of Machine Learning Research}}, vol. 21, no. 1, pp. 5485-5551, 2020.

\bibitem{ref9} J. Devlin, M. W. Chang, K. Lee, and K. Toutanova, \textquotedblleft BERT: Pre-training of deep bidirectional transformers for language understanding," {\it{arXiv}} preprint arXiv:1810.04805, 2018.

\bibitem{ref10} E. Aghaei, X. Niu, W. Shadid, and E. Al-Shaer,  \textquotedblleft SecureBERT: A Domain-Specific Language Model for Cybersecurity," in {\it{ International Conference on Security and Privacy in Communication Systems}}, pp. 39-56, Cham: Springer Nature Switzerland, 2022.

\bibitem{ref11} H. Touvron, L. Martin, K. Stone, P. Albert, A. Almahairi, Y. Babaei, N. Bashlykov, \textquotedblleft Llama 2: Open foundation and fine-tuned chat models," {\it{arXiv}} preprint arXiv:2307.09288, 2023.

\bibitem{ref12} L. Yang, A. Moubayed, and A. Shami, \textquotedblleft MTH-IDS: A multitiered hybrid intrusion detection system for internet of vehicles," {\it{IEEE Internet of Things Journal}}, vol. 9, no. 1, pp. 616-632, 2021.



\bibitem{ref13} H. M. Song, J. Woo, and H. K. Kim, \textquotedblleft In-vehicle network intrusion detection using deep convolutional neural network," {\it{Vehicular Communications}}, vol. 21, 2020.

\bibitem{ref14} X. Li and H. Fu, \textquotedblleft A Hybrid Ensemble Multilayer-Perceptron-based Intrusion Detection System for Vehicle Networks," in {\it{2023 International Institute of Cognitive Informatics and Cognitive Computing, IEEE}}, 2023.

\bibitem{ref15} E. Seo, H. M. Song, and H. K. Kim, \textquotedblleft GIDS: GAN based intrusion detection system for in-vehicle network," in {\it{2018 16th Annual Conference on Privacy, Security and Trust (PST)}}, pp. 1-6, IEEE, 2018.

\bibitem{ref16} T. P. Nguyen, H. Nam, and D. Kim, \textquotedblleft Transformer-Based Attention Network for In-Vehicle Intrusion Detection," {\it{IEEE Access,}} 2023.

\bibitem{ref17} A. Alshammari, M. A. Zohdy, D. Debnath, and G. Corser, \textquotedblleft Classification approach for intrusion detection in vehicle systems," {\it{Wireless Engineering and Technology}}, vol. 9, no. 4, pp. 79-94, 2018.

\bibitem{ref18} V. S. Barletta, D. Caivano, A. Nannavecchia, and M. Scalera, \textquotedblleft A Kohonen SOM architecture for intrusion detection on in-vehicle communication networks," {\it{Applied Sciences}}, vol. 10, no. 15, 2020.

\bibitem{ref19} H. Olufowobi, C. Young, J. Zambreno, and G. Bloom, \textquotedblleft Saiducant: Specification-based automotive intrusion detection using controller area network (CAN) timing," {\it{IEEE Transactions on Vehicular Technology}}, vol. 69, no. 2, pp. 1484-1494, 2019.

\bibitem{ref20} S. F. Lokman, A. T. Othman, M. H. Abu Bakar, and R. Razuwan, \textquotedblleft Stacked sparse autoencodersbased outlier discovery for in-vehicle controller area network (CAN)," {\it{Int. J. Eng. Technol}}, vol. 7, no. 4.33, pp. 375-380, 2018.

\bibitem{ref21} J. Ashraf, A. D. Bakhshi, N. Moustafa, H. Khurshid, A. Javed, and A. Beheshti,  \textquotedblleft Novel deep learning-enabled LSTM autoencoder architecture for discovering anomalous events from intelligent transportation systems," {\it{IEEE Transactions on Intelligent Transportation Systems}}, vol. 22, no. 7, pp. 4507-4518, 2020.


\bibitem{ref22} E. J. Hu, Y. Shen, P. Wallis, Z. Allen-Zhu, Y. Li, S. Wang, L. Wang, and W. Chen, \textquotedblleft Lora: Low-rank adaptation of large language models," {\it{arXiv}} preprint arXiv:2106.09685, 2021.


\bibitem{ref23} E. Nwafor and H. Olufowobi,  \textquotedblleft CANBERT: A Language-based Intrusion Detection Model for In-vehicle Networks," in {\it{ 2022 21st IEEE International Conference on Machine Learning and Applications (ICMLA)}}, pp. 294-299, IEEE, 2022.

\bibitem{ref24} N. Alkhatib, M. Mushtaq, H. Ghauch, and J. L. Danger, \textquotedblleft  CAN-BERT do it? Controller Area Network Intrusion Detection System based on BERT Language Model," in {\it{2022 IEEE/ACS 19th International Conference on Computer Systems and Applications (AICCSA)}}, pp. 1-8, IEEE, 2022.

\bibitem{ref25} Y. Zhu, R. Kiros, R. Zemel, R. Salakhutdinov, R. Urtasun, A. Torralba, and S. Fidler,\textquotedblleft Aligning books and movies: Towards story-like visual explanations by watching movies and reading books," in {\it{Proceedings of the IEEE International Conference on Computer Vision}}, pp. 19-27, 2015.

\bibitem{ref26} I. Loshchilov and F. Hutter,  \textquotedblleft Decoupled weight decay regularization," {\it{arXiv}} preprint arXiv:1711.05101, 2017.

\bibitem{ref27} S. Mangrulkar, \textquotedblleft PEFT: State-of-the-art Parameter-Efficient Fine-Tuning methods," [Online]. Available: https://github.com/huggingface/peft, 2022.

\bibitem{ref28} F. Salo, M. Injadat, A. B. Nassif, and A. Essex, \textquotedblleft Data mining with big data in intrusion detection systems: A systematic literature review," {\it{arXiv}} preprint arXiv:2005.12267, 2020.




\end{thebibliography}
\end{document}